\documentclass{emulateapj}
\usepackage{graphicx}
\usepackage{amssymb}
\usepackage{epstopdf}
\usepackage{wasysym}
\usepackage{natbib}

\shorttitle{Early Environment of Dwarf Galaxies}
\shortauthors{Corlies et al.}

\begin{document}
\title{Chemical Abundance Patterns and the Early Environment of Dwarf Galaxies}
\author{Lauren Corlies,\altaffilmark{1} Kathryn V. Johnston,\altaffilmark{1} Jason Tumlinson,\altaffilmark{2}  Greg Bryan\altaffilmark{1}}

\altaffiltext{1}{Department of Astronomy, Columbia University, New York, NY 10027, USA; lauren@astro.columbia.edu}
\altaffiltext{2}{Space Telescope Science Institute, Baltimore, MD 21218, USA}

\begin{abstract}
Recent observations suggest that abundance pattern differences exist between low metallicity stars in the Milky Way stellar halo and those in the  dwarf satellite galaxies. This paper takes a first look at what role the early environment for pre-galactic star formation might have played in shaping these stellar populations. In particular, we consider whether differences in cross-pollution between the progenitors of the stellar halo and the satellites could help to explain the differences in abundance patterns. Using an N-body simulation, we find that the progenitor halos of the main halo are primarily clustered together at $z=10$ while the progenitors of the satellite galaxies remain on the outskirts of this cluster. Next, analytically modeled supernova-driven winds show that main halo progenitors cross-pollute each other more effectively while satellite galaxy progenitors remain more isolated.  Thus, inhomogeneous cross-pollution as a result of different high-$z$ spatial locations of each system's progenitors can help to explain observed differences in abundance patterns today. Conversely, these differences are a \emph{signature} of the inhomogeneity of metal enrichment at early times.
\end{abstract}

\section{Introduction}
For several decades now, hierarchical formation of dark matter halos, or small halos merging to form larger halos, has provided the framework for theories of galaxy formation.  Within this paradigm, the stellar halo and substructures within it (such as satellites and stellar streams) can arise naturally from the accretion and subsequent disruption of dwarf galaxies \citep{BullockJohnston}.  A potentially serious challenge to this theory emerged when observations revealed systematic differences in abundances between stars in the classical dwarf spheroidals and those in the stellar halo:  at  [Fe/H]$\approx$-2, halo stars were found to have higher [$\alpha$/Fe] than those in the dwarf galaxies \citep[see summary by][]{venn04}.  If the stellar halo was built up from dwarf galaxies, how could these abundance differences exist?  However, it was soon realized that the infalling objects that build the bulk of the stellar halo would have been accreted predominately at early times before the explosion of Type Ia supernovae, when only  high-$\alpha$ stars would have had time to form.  In contrast, the current dwarf spheroidal satellites would have typically been accreted at later times, allowing for some low-$\alpha$ stars to form from gas which had been polluted by Type Ia supernovae, which could account for the abundances differences \citep{robertson05,font06}.  

More recently, further paradoxes in abundance patterns have emerged when examining lower-metallicity stars ([Fe/H] $\approx$ -3): neutron capture elements such as Barium and Strontium were found to have systematically lower abundances in the Ultrafaint dwarf satellite galaxies (UFDs) than in the Milky Way stellar halo \citep[see compilation by][]{frebel10};  and $\alpha$-element abundances appeared to be higher in the classical dwarf satellites as opposed to the stellar halo \citep{kirby11}.  If it is assumed that the {\it progenitors} of the disrupted and surviving satellite galaxies are the same except for their accretion times, then the low-metallicity stars in both systems  might be expected to have the same abundance patterns and any differences again seem to challenge the hierarchical structure formation paradigm. Alternatively, the disparity could be used to inform one about the nature of the progenitors.  The dwarfs that form the satellite system of the Milky Way are potentially different from those that formed the stellar halo by virtue of the fact that they were accreted later and survived.  The disparity in accretion times in turn suggest other possible differences between satellite and halo progenitors in the early Universe, in particular between their spatial locations and hence environment. Indeed, the nature of star formation in the first galaxies is expected to be heavily influenced by patchy re-ionization, inhomogeneous chemical enrichment from Population III stars and cross-pollution from neighboring systems. Thus, these abundance pattern differences can potentially serve as a unique window on these early times and speak to the very way in which the Galaxy formed within the hierarchical model.

Abundance patterns are dictated by several factors. For example, larger systems are typically influenced by pollution from a more fully sampled stellar initial mass function (IMF).  Thus, the effective yield from a combination of many different mass supernovae produces an averaged abundance pattern in the stars with small dispersion.  Conversely, lower mass galaxies may be polluted by a more incompletely sampled stellar IMF and can be influenced by stochastic effects.  In particular, \citet[][submitted]{Duane_paper} showed the skew of the distribution of neutron capture elements in the UFDs relative to the distribution seen in the stellar halo could arise from the stars in the smaller objects being polluted by the highly-mass dependent yields of a small number of supernovae.  Note, however, that this picture is only strictly true for galaxies evolving in isolation.  If low mass galaxies happen to experience high levels of cross-pollution from their neighbors, the abundance patterns would then reflect the effective yield of many supernovae, leading to the same average value we would expect for more massive galaxies.  Hence, it is the combination of galaxy mass and environment that determine the abundance patterns.

With this understanding, the measured abundance differences between halo stars and the UFDs can naturally be explained if one of these classes of objects is preferentially
\begin{description}
\item{\it polluted} ---  if the first stars have unique yields as is currently assumed, differences in the amount of pollution from these stars could be driving the abundance differences
\item{\it isolated} ---  if the dwarf galaxy progenitors are chemically isolated, their abundance patterns would reflect stochastic pollution, while the cross-polluting main halo progenitors would converge to an average
\end{description}
Both of these explanations require differences among the level of cross-pollution (and similarly isolation) experienced by these two classes of objects. This naturally leads to the main questions that are the focus of this work: what is the role of cross-pollution and does it affect the progenitors differently?  To what extent are the progenitors of the different classes of objects (i.e. satellites vs. the main halo) isolated at high $z$? 

The most obvious way to address these questions is using detailed, cosmological, hydrodynamical simulations of the early formation and environment of a massive galaxy and its satellites.  Such a simulation would need to track individual sub-galaxies forming in a Milky Way-like main halo at high redshift to the present day with high enough resolution to accurately follow inhomogeneous pollution from these halos and the effects on subsequent star formation \citep[e.g. work by][suggests a resolution of better than 10 pc is needed to follow outflows from low-mass dwarf galaxies] {simpson13}.  State-of-the-art simulations are not yet able to meet these combined criteria.  For example, \citet{john_wise} followed the cross-pollution of dwarf galaxies in both dense environments and voids with a maximal resolution of 1 pc comoving using Enzo \citep{enzo}.  However, to meet these specifications, the simulation was only run to $z=7$ instead of to $z=0$ in a 1 Mpc box, which is too small to encompass the progenitors of a Milky Way-like halo.  In comparison, \citet{zolotov_paper} is an example of cutting edge simulations of a Milky Way-sized halo run to $z=0$ using GASOLINE \citep{gasoline}.  Their simulations included stellar feedback as well as metal enrichment, cooling and diffusion.  However, computational cost dictated that these simulations were of lower resolution than those of \citet{john_wise}, which compromises their ability to accurately follow outflows and metal mixing on the small scales of interst here.  These two works demonstrate the difficulties involved in tracing the effects of high-$z$ environment and star formation until the present day in a single simulation for both Eulerian and Lagrangian codes.  Moreover, neither type of code is yet able to follow all the relevant physics directly but rather rely on sub-grid models that are calibrated to observations.

Our own interest in examining how differences in high-$z$ environment affect present day abundance signatures in dwarf galaxies lies at the intersection of these two simulations.  Higher resolution than \citet{zolotov_paper} is required to trace low-mass progenitors at early times and larger volumes than \citet{john_wise} run to present day are needed to examine a Milky Way-like galaxy.  Thus, to motivate the intense computational costs of such a simulation, this paper takes a first look at how high-$z$ cross-pollution varies within a Milky Way-like halo using more simplified techniques.  In section 2, we use an N-body simulation to trace where objects identified today came from in the early universe and whether these high-$z$ spatial locations can vary systematically with progenitor type.  In section 3, we use an analytical model of supernova-driven winds to estimate if the importance of cross-pollution might differ across the progenitors of different objects.  Finally, our conclusions are presented in section 4.

\section{Results I: Origin of Progenitors}

To begin to address the role of early environment in shaping stellar populations, we first identify the origins of the progenitors of the current main halo and dwarf galaxies.  In particular, we are interested in investigating if there are systematic differences in spatial location of those progenitors at \emph{high-$z$} that reflect systematic differences in spatial location at \emph{low-$z$} of the main halo and its satellites and if these differences have observable consequences.   This is done using an N-body simulation described below.

This approach is inspired by \citet{Diemand_sigmapeaks} who showed that early ($z \gtrsim$ 10) high-$\sigma$ peaks in density fluctuations influenced the present-day spatial distribution and kinematics of protogalactic systems.  They found that material associated with these rarer peaks was more centrally concentrated in their present halos and moved with a lower velocity dispersion on more radial orbits.  Here, a similar question is asked but in the reverse sense.  Instead of tracing structures forward in time, we ask: where do the structures seen today come from?; and is there a systematic difference in where objects such as dwarf galaxies form as opposed to where the main halo forms?

\subsection{Method: N-Body Simulation}

The N-body simulation that is the underpinning of this work is described by \citet{Jason_simulation}.  A Milky Way-like halo was identified from within a full cosmological simulation of a periodic cubic box of comoving size 7.320$h^{-1}$ Mpc in one dimension using Gadget2. \citep[ver 2.0;][]{gadget2}    The main halo was found to have a virial mass of $M_{200}=1.63 \times 10^{12} M_{\astrosun}$ and a virial radius of $R_{200}=381$kpc.  The Milky Way-like halo was then re-simulated with the central portion of the galaxy having a higher resolution (512$^3$) while the rest of the box was run at a lower resolution (256$^3$) to save time.  At the 512$^3$ resolution, the dark matter particles had a mass of $M_p = 2.64 \times 10^5 M_{\astrosun}$.  The gravitational smoothing length for all simulations was 100 pc in comoving coordinates.  Subsequently, the six-dimensional friends-of-friends algorithm \citep{FOFpaper} was implemented to identify all bound halos with a mass of at least $8 \times 10^6$ M$_{\astrosun}$ at each time step in the simulation.  The algorithm identified the main halos as well as the small halos embedded within their larger potential well at later times. 

Furthermore, a semi-analytic star formation history was prescribed for each halo, which was then used to identify luminous satellites at z=0.  This star formation history included chemical and kinematic feedback as well as reionization.  Details can be found in \citet{Jason_simulation} Section 4.  The prescription reasonably matched observed traits of the Milky Way and its satellite populations.  For the simulation's main halos, both the overall stellar mass and the stellar mass density profiles were well reproduced, indicating that the chosen parameter values of the basic baryon assignment and star formation prescription were well chosen.  In addition, the prescription matched both the luminosity function and the metallicity distribution function of the Milky Way satellites, demonstrating that the implemented feedback was accurate down to the smallest mass scales of interest.  For these reasons, this work assumes all the fiducial values of \citet{Jason_simulation} in our own models with high confidence.

\subsection{Main Halo and Dwarf Progenitor Clustering}

The aim is to compare the early environment of main halo progenitors with that of dwarf galaxy progenitors by analyzing two snapshots from the full simulation described above.  Figure 1 shows the $z=0$ snapshot.  Particles within the inner 25 kpc of the main halo (hereafter ``main halo particles'') are shown in blue.  \citep[Note that the cost of high-resolution spectroscopy means that abundance distributions for the stellar halo have typically been derived from even more local stars, so including halo particles at greater distance from the center would not fairly represent the observed samples --- see, e.g.,][]{venn04}.
Dwarf galaxies are defined as surviving, non-dark subhalos at $z=0$ \citep[i.e. those that are assumed to contain stars using the semi-analytic prescriptions of][]{Jason_simulation} --- 124 surrounding halos are identified as dwarf galaxies from the previously constructed halo catalogues. These ``dwarf galaxy particles'' are shown in red. 

\begin{figure}
\begin{center}
\includegraphics[width=0.5\textwidth]{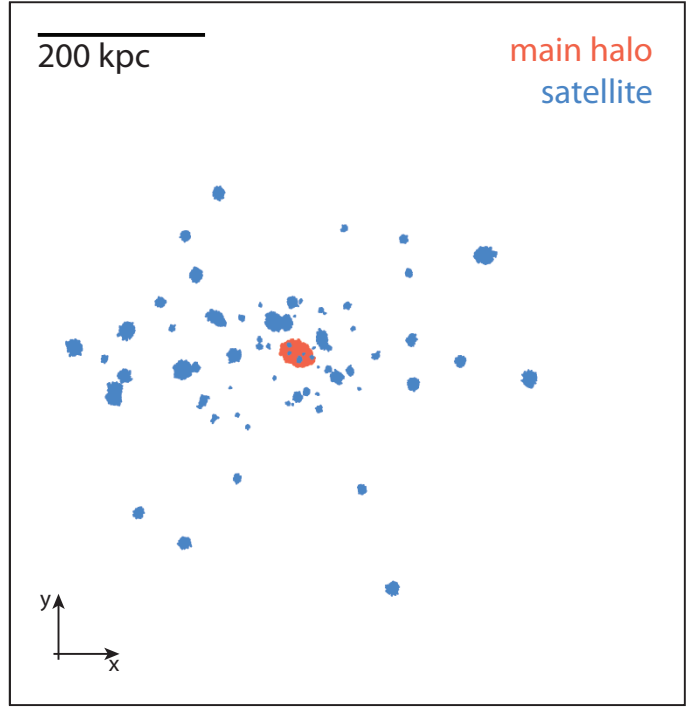}
\caption{ Plotted is x-y projection of the $z=0$ snapshot of the simulation. The main halo sits in the center (red) and is surrounded by the dwarf galaxies (blue) \label{z0.fig}}
\end{center}
\end{figure}

\begin{figure*}
\begin{center}
\includegraphics[scale=0.75]{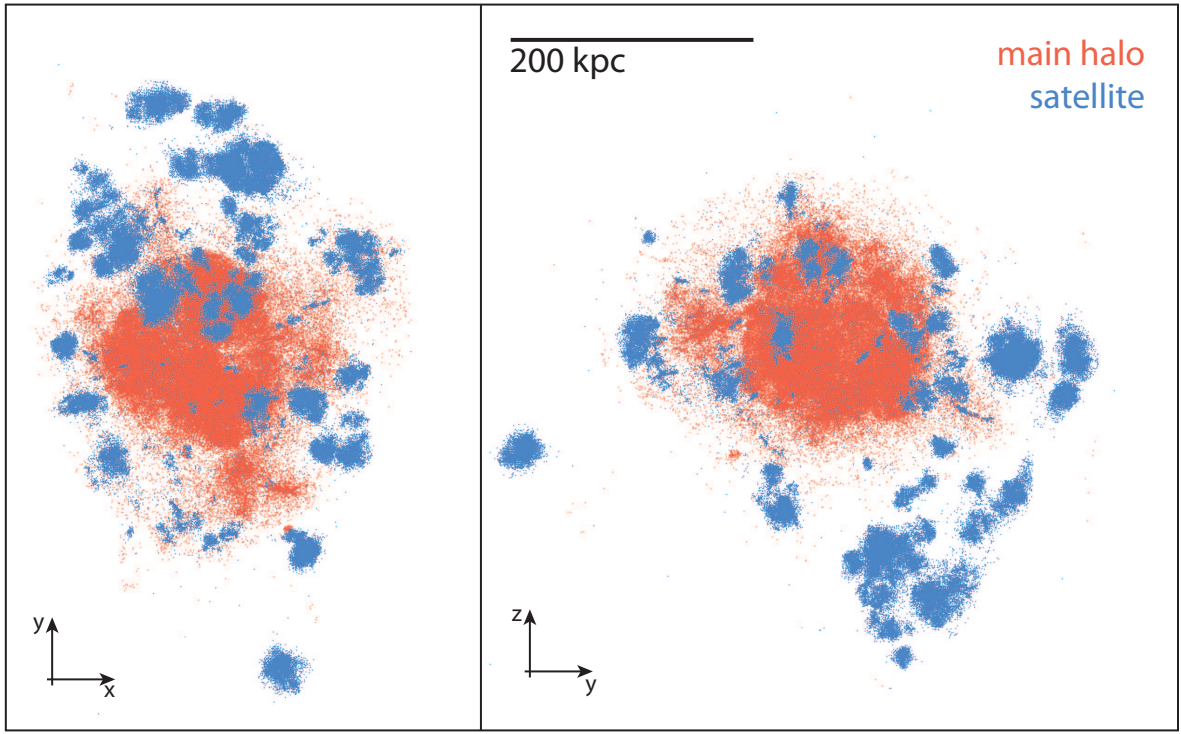}
\caption{ Plotted are x-y (left) and y-z (right) projections of the $z=10$ positions of the particles that comprise the main halo (red) and dwarf galaxies (blue) at $z=0$.  Bound halos at $z=10$ with mass greater than 10$^8$ M$_{\astrosun}$ that contain any of these particles are overplotted as a square for main halo progenitors and as a star for dwarf galaxy progenitors. \label{projections.fig}}
\end{center}
\end{figure*}

Figure 2 shows two projections of the second snapshot at $ z=10$. The same sets of particles identified at $z=0$ (halo in red, satellites in blue) are shown to indicate their formation environment.  This redshift is the redshift of reionization in the full model, at which point star formation is truncated in the dwarf galaxies and smaller main halo progenitors \citep[see][]{Jason_simulation}, making it a logical choice for studying the early environment.  At such early times, more than 80$\%$ of eventual main halo and dwarf galaxy particles are still not bound to any progenitor halo.  As an example, overplotted on the particles are all bound halos that contain any of these particles with mass greater than 10$^8 $M$_{\astrosun}$.  Main halo progenitors are represented as squares and the dwarf galaxy progenitors are represented as stars in Figure \ref{projections.fig}.  Visual inspection of these projections is enough to demonstrate that one cannot simply assume that the formation environments of dwarf galaxy progenitors and main halo progenitors are the same.  Instead, dwarf galaxy progenitor particles appear to sit preferentially outside of the main halo progenitor particles, which are clustered in the center of the formation region.  Expressed more quantitatively, 50$\%$ of the main halo particles can be enclosed in a physical radius of 60 kpc while a physical radius of 150 kpc is needed to enclose 50$\%$ of the dwarf galaxy particles.  Making similar plots for three additional Milky Way-like galaxies displayed the same relations.  Thus, this spatial separation of the two populations of particles appears to be a general result of hierarchical structure formation.

We conclude that the dwarf galaxy progenitors and main halo progenitors have distinctly different spatial locations at $z=10$.  However, the spatial isolation of the dwarf galaxy progenitors does not ensure their chemical isolation.  To determine that these progenitors are also chemically isolated, we now examine pollution due to supernova-driven winds.

\section{Results II: Cross-Pollution}
Winds driven by Type II supernovae (SNe II) are considered to be one of the main mechanisms for enriching the intergalactic medium (IGM) with metals \citep{Greg_snpollution}.  It follows that inhomogeneous pollution from these winds, especially in the form of preferential cross-pollution of certain halos, could have large effects on the abundance patterns of stars forming in different regions of space.  Given the differences in spatial locations of main halo and dwarf galaxy progenitor particles observed in our simulations, one could expect to see differences in cross-pollution across these two classes of progenitors. 

To follow cross-pollution, we build on previous work that semi-analytically calculated radii of SN-driven bubbles.  Our predecessors were interested in examining these SNe as a way to drive reionization \citep{Tegmark_winds} as well as looking at how they advanced metal enrichment to better understand metal absorption lines \citep{Furlanetto_winds} and the locations of the first stars \citep{Scannapieco_winds}.  These works were focused on global enrichment or reionization of the IGM by galaxy populations, and so averaged over many halos in large volumes and considered longer timescales.  

While the scheme used is similar to previous work, there is a significantly different motivation here.  Instead of examining more global processes, a single forming galaxy is analyzed for a short part of its early history.  Furthermore, we have the advantage of using a full N-body simulation as the spatial basis.  This makes it possible to look at how the winds are spatially distributed on these smaller scales as well as the level of cross-pollution among the halos.

In order to investigate cross-pollution, we implement a basic SN-driven wind model as in these previous works and apply it to individual halos from the N-body simulation.  The method is described in the section 3.1 while the results are presented in section 3.2.

\subsection{Bubble Evolution}

The SN-driven winds are modeled analytically as spherically symmetric, thin shells in an expanding universe (with density parameters  for dark energy, dark matter and baryons of $\Omega_{\Lambda}=0.762, \Omega_M=0.238, \Omega_b=0.0416$ respectively and Hubble constant, $H_0=73.2 $\ km s$^{-1}$ Mpc$^{-1}$) of zero pressure and constant IGM density, $\bar{\rho}$.  The shell is assumed to sweep up almost all of the baryonic IGM that it encounters.  With these assumptions, the evolution of the shell radius, R$_{s}$, is given by:
\begin{equation}
    \ddot{R}_s = \frac{3P_b}{\bar{\rho}R_s} - \frac{3}{R_s}(\dot{R}_s - HR_s)^2 - \Omega_m \frac{H^2R_s}{2}
\end{equation}
where the overdots represent time derivatives and the subscripts $s$ and $b$ indicate shell and bubble quantities respectively. The first term in Equation 1 describes how the interior pressure, $P_b$, drives the bubble expansion.  The bubble is slowed by the fact that newly swept up mass must be accelerated from its Hubble flow velocity, $HR_s$, to $\dot{R_s}$ so that the second term in Equation 1 represents a net braking force.  Finally, the third term represents the gravitational pull from the mass interior to the shell which again decelerates the shell. 

The pressure, $P_b$, that drives the bubble is provided by the SNe, which have a net input of energy into the system with a rate equal to
\begin{equation}
	\dot{E}_b = L(t) - 4\pi R^2_s \dot{R}_sP_b.
\end{equation}
Here, $L(t)$ is the luminosity of the SNe and the remaining term is the typical work done by the shell as it expands.  Lastly, adiabatic expansion is assumed such that $P_b = E_b/2 \pi R^3_s$.

The only unique function that can be specified in the above equations is the SNe luminosity as defined by the halo's star formation history and its mass accretion history. Prescriptions for both mass accretion and star formation are implemented and are described below.

\subsubsection{Mass Accretion Histories}
The mass accretion history of each halo is calculated using the formalism of \citet{Wechsler_masshistories}.  After examining a large range of full mass assembly histories from high resolution N-body simulations, they found that the following simple parameterization captured most crucial aspects of the  growth in the mass of a halo, $M_{\rm halo}$, over time:

 \begin{equation}
	M_{\rm halo}(z) = M_0 \exp \left[\frac{-S}{1+z_c} \left(\frac{1+z}{1+z_0}-1 \right) \right]
\end{equation}
where S=2.0, $z_0$ was the observing redshift and $M_0$ was the observed mass at that redshift. The characteristic formation time, $z_c$  was defined to be:

 \begin{equation}
 	1 + z_c = \frac{c_v}{c_1} (1+z_0)
 \end{equation}
where $c_1$=4.1 was the typical concentration of a halo forming at $z=0$ and $c_{\rm{vir}}$ was the concentration of the halo in question.
 
Individual halos can deviate significantly from this simple form, especially around a major merger.  However, using this type of smooth, continuous accretion history instead of the actual simulated histories is justified as there is not expected to be many major mergers at this early time or during such a short time period.  Furthermore, this semi-analytic parameterization was chosen over the full merger history for each halo because it was straightforward to incorporate into the wind model without much loss of accuracy.

For this study, the N-body simulation is used to find the virial radius and half-mass radius of the subhalos identified by the group finder at $z=10$ that contain eventual main halo and dwarf galaxy particles. We can then calculate the concentration of each halo for which we assign a mass accretion history.

\subsubsection{Star Formation Histories and Supernovae Luminosity}
	The final step in defining the SN luminosity is prescribing a star formation history for each halo.  A simplified version of the prescription applied in \citet{Jason_simulation} is used. 
	Dark matter halo masses, from Equation 3 are converted to baryonic gas masses with a fixed efficiency prior to reionization:
	\begin{equation}
		M_{\rm{gas}} = f_{\rm{bary}} \times M_{\rm{halo}}
	\end{equation}
where f$_{\mathrm{bary}}$ = 0.05.

In any time interval, $\Delta t$, the mass of stars formed is:
	\begin{equation}
		M_{\rm{star}} = \frac{M_{\rm{gas}} \Delta t}{\tau_{\rm{sf}}}
	\end{equation}
where $\tau_{\rm sf}$ = 10 Gyr.
Finally, following \citet{Jason_simulation}, it is assumed that 1 M$_{\astrosun}$ of star formation yields 0.01 SNe for a \citet[][]{kroupa_imf} IMF and that each supernova produces 10$^{51}$ ergs of energy.  This comes from integrating the IMF for the number of stars between 8-40 M$_{\astrosun}$ which yield such supernovae and normalizing to 1 M$_{\astrosun}$. 
Thus, the SNe luminosity can be defined as:
	\begin{equation}
		L = (0.01 \ \mathrm{SNe/M}_{\astrosun}) (10^{51} \mathrm{ergs/SNe}) \ \frac{f_{\rm{bary}} M_{\rm{halo}}}{\tau_{\rm{sf}}}
	\end{equation}

This SNe luminosity differs significantly from the previous works cited above \citep{Tegmark_winds, Scannapieco_winds, Furlanetto_winds}.  They chose to have a single starburst per galaxy while we allow for more realistic, continuous star formation.  Furthermore, dissipation effects are not included, such as the ionization of the swept-up IGM or heating of the shell and interior plasma. Thus, these estimates give an upper limit for the wind radius and maximize the effects of cross-pollution. (In general, this choice is insignificant because of the short integration time, and the radius changes by at most 2$\%$ when dissipation is included.)  Because the principle objective of this work is to explain abundance pattern differences in stars with [Fe/H] $\approx$ -3, the prior presence of Population III stars in the galaxies is also not modeled.  These stars will simply provide the initial enrichment necessary for star formation at the metallicities considered here.

Finally, a treatment of reionization is not included because the process is still not well characterized on these small scales.  In reality, each of the supernova-driven bubbles would have been preceded by a reionization front, potentially limiting further star formation in neighboring halos.  However, the importance of inhomogeneous reionization {\it within} a Milky Way-like galaxy is only just being studied \citep[e.g.][]{ocvirk11}, with most previous work concentrating on global (and typically external) influences on the entire satellite system \citep{busha10,lunan12}.  Furthermore, any simple analytic treatment of reionization is dependent on the value of the escape fraction of the ionizing photons, which is currently not well known.  Thus, rather than using a poorly constrained reionization model, we point out that including reionzation can only strengthen these results as reducing or even truncating star formation in neighboring systems can only lessen the bubble radii and overlap.

\subsubsection{Integration Time}
The final piece needed to compute the supernova-driven wind radii is the time  over which to integrate the star formation histories.  This time frame needs to result in stars with a metallicity consistent with those of the low-metallicity stars whose abundance patterns this paper seeks to explain.  In order to make an order-of-magnitude estimate of a halo's metallicity, we use a simple closed box model with all metals produced by the SNe remaining in the host halo \citep[as described in][]{Binney_book}. Although a mass accretion history is assumed (Section 3.1.1), this closed box assumption is valid because over the short time considered here, a typical 10$^8 $ M$_{\astrosun}$ halo accretes less than $2\%$ more dark matter and gas.  Furthermore, it is assumed that all the metals ejected by the supernovae are evenly and instantaneously mixed throughout the entire bubble.  Using the star formation history described in Section 3.1.2, it is then straightforward to show that the closed-box model becomes:
\begin{equation}
Z(t) = -p \ln \left [1 -  \frac{t}{\tau_{\rm{sf}}} \right ].
\end{equation}

\begin{figure*}
\epsscale{0.85}
\plotone{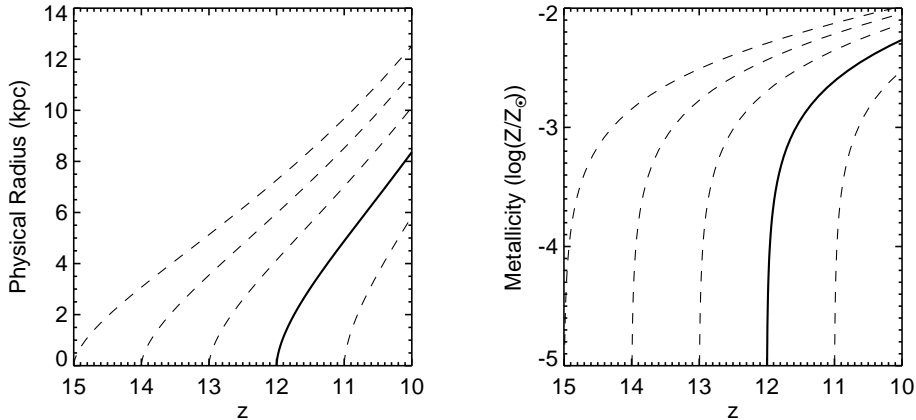}
\caption{In the left-hand panel, the physical radius of the SNe-driven wind is shown as a function of redshift.  In the right-hand panel, the metallicity is plotted as log(Z/Z$_{\astrosun}$) for a $10^8$ M$_{\astrosun}$ halo as a function of redshift.  In the right, the mass in stars as a function of redshift is shown. The metallicity directly reflects the amount of gas that is being turned into stars. \label{metal_evol.fig}  The final values of each quantity do not vary much as a function of the starting redshift.  The solid lines indicate a starting redshift of $z=12$ and are the values adopted henceforth.}
\end{figure*}

Thus, because of the simplicity of this model, a star formation rate that is proportional to the dark matter mass and an identical start time for star formation all halos, the metallicity depends only on the time passed since the beginning of star formation.  A single metallicity value for all of our halos is produced.  The metallicity is plotted as log(Z/Z$_{\astrosun}$) for a $10^8$ M$_{\astrosun}$ halo in the right-hand panel of Figure \ref{metal_evol.fig} as a function of redshift for a variety of start times.  Overall, the plot suggests that these halos should be forming stars at $z=10$ with an approximate metallicity of log(Z/Z$_{\astrosun}$)= -2.25, regardless of when star formation begins.  This simple estimate is consistent with time and mass dependent chemical evolution models from \citet{Jason_simulation}, for these same simulated halos.  This demonstrates that the metallicity of the stars at this time are expected to be in the range of the low metallicity stars with observed abundance differences.

\subsection{Findings}

Physical bubble radii for every progenitor halo are calculated by beginning star formation at $z=12$ for all progenitor halos found at $z=10$, with the results being fairly insensitive to this choice in starting redshift.  In the left-hand panel of Figure \ref{metal_evol.fig}, the radius is plotted as a function of $z$ for an example halo taken from the simulation with a mass of 10$^8$ M$_{\astrosun}$ and a measured concentration of 6.57.  This shows that the scale of the radii is roughly consistent with the work of 
\citet{Tegmark_winds}, \citet{Scannapieco_winds}, and \citet{Furlanetto_winds} with the differences arising from the differences in our luminosity function, described above.
For completeness, the radius evolution is also shown for a number of starting redshifts extending to $z=15.$  While the choice in starting redshift does have an effect on the final radius at $z=10$, the change in radius is small enough that it does not affect our conclusions concerning cross-pollution.

\begin{figure}
\begin{center}
\includegraphics[width=0.40\textwidth]{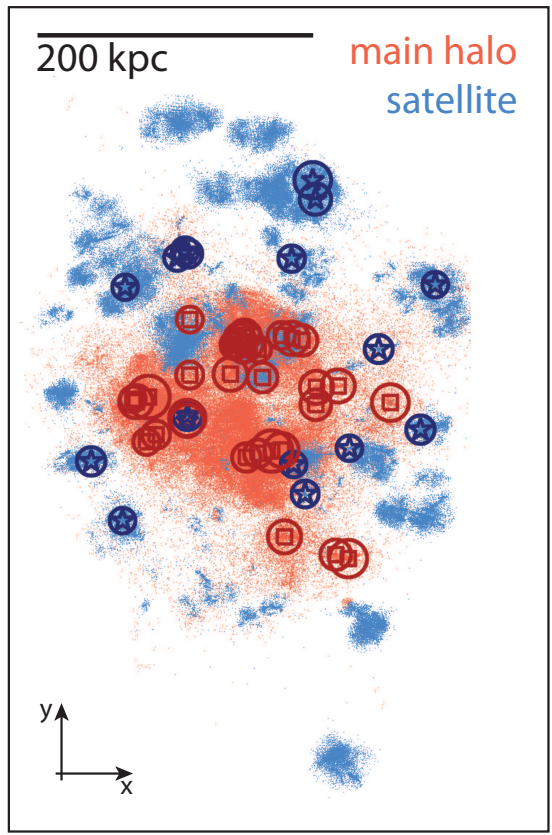}
\caption{Figure 1 is again presented but now with the shells of the SN-driven winds overplotted with main halo progenitors in red and dwarf galaxy progenitors in blue.  For simplicity, only those bound halos with mass greater than 10$^8$M$_{\astrosun}$ and their radii are shown.  The main halo progenitors (squares) are more centrally located with an apparently higher cross-pollution rate while the dwarf galaxy progenitors (stars) remain more isolated on the outskirts. \label{bubble_overplot.fig}}
\end{center}
\end{figure}

The order of magnitude of the bubble radii also demonstrates that the halos have the ability to pollute a large surrounding volume, even at this early time.  For a sense of scale, Figure \ref{bubble_overplot.fig}, which plotted main halo (red) and dwarf galaxy (blue) progenitor particles, is shown again with the bubble radius overplotted for each halo.  To most clearly demonstrate the result, only bound halos with mass greater than 10$^8$ M$_{\astrosun}$ are included again.  This projection indicates that the bubbles of main halo progenitors overlap more often than their dwarf counterparts.  Thus, one might expect that because of their clustering in the center, main halo progenitors would have a higher level of cross-pollution while the dwarf galaxy progenitors would remain more isolated.  This type of non-uniform pollution is easily seen in \citet{john_wise} (Figure 1).  Closely clustered halos are much more capable of spreading their metals amongst themselves within their simulation and this pollution is seen to affect how star formation proceeds in such instances.  Because of computational limitations, these halos are unable to be identified as either main halo progenitors or dwarf galaxy progenitors in the manner described here but the effects of the cross-pollution are meaningful.

\begin{figure*}
\epsscale{0.85}
\plotone{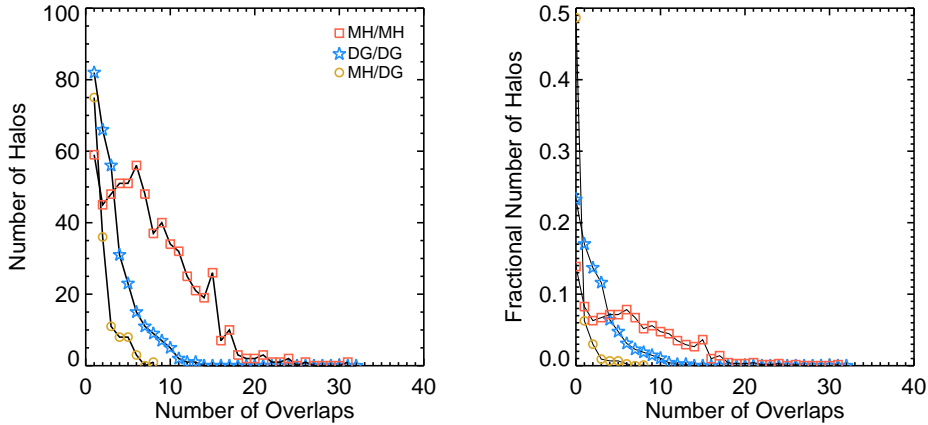}
\caption{The left panel demonstrates that the bubbles of main halo progenitors (squares) overlap more often than those of their dwarf galaxy counterparts (stars). Cross-pollution among main halo and dwarf galaxy progenitors (circles) is rarer and typically only involves one crossing. The right panel, plotting the fraction of halos per crossing, shows that this trend is not simply due to the larger number of main halo progenitors. \label{overlap_histo.fig}}
\end{figure*}

The left panel of Figure \ref{overlap_histo.fig} illustrates the significance of cross-pollution and this potential pollution bias by showing the number of times the bubbles of the different classes of progenitors cross.  The bubbles were defined as crossing if the distance between the centers of the two halos was less than the sum of their radii.  It shows that main halo progenitors tend to cross-pollute other main halo progenitors (squares) much more often than dwarf galaxy progenitors cross-pollute other dwarf galaxy progenitors (stars).  Moreover, main halo and dwarf galaxy progenitors (circles) rarely cross-pollute each other.  The right-hand panel plots for each crossing category, the fraction of halos that experience a given number of pollution crossings.  This demonstrates that the effect is not simply due to the fact that there are a larger number of main halo progenitors.

\begin{figure}
\begin{center}
\includegraphics[width=0.37\textwidth]{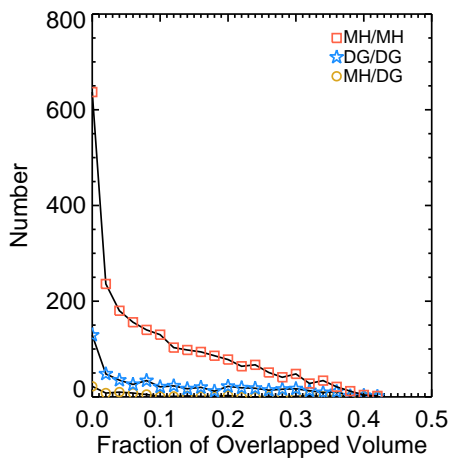}
\caption{The fraction of volume overlap for each pair of bubbles is illustrated.  Not only do main halo progenitors (squares) cross-pollute each other more often but the generally higher volume overlap indicates that they also do so more effectively.  \label{volume_overlap.fig}}
\end{center}
\end{figure}

Because actual pollution is not expected to be exactly spherical, simply examining whether bubbles cross or not might not represent true cross-pollution.  Thus, a more significant calculation is what fraction of the two bubbles' volumes overlap.   This is defined as the volume of overlap divided by the sum of the two individual bubble volumes.  Bubbles that have a large fractional volume of overlap have a much higher probability of cross-pollution regardless of potential asymmetries in the bubble structure.  For simplicity, we calculate the fractional volume of each pairwise overlap without considering multiple bubble overlaps.  Figure \ref{volume_overlap.fig} indicates that the fraction of overlapping volume ranges from a small fraction to as high as 40$\%$ for cross-pollution among main halo progenitors (red) and dwarf galaxy progenitors (blue).  Thus, with similar volume distributions, the higher number of main halo crossings truly represents a higher level of cross-pollution.  On the other hand, the majority of main halo - dwarf galaxy cross-pollution events involve less than 10$\%$ of the volume involved.  Thus, the spread in the fractional volume overlap reinforces how spatially dependent the process of cross-pollution is.

However, because such early times are examined here, many of the dwarf galaxy progenitors will merge to form larger dwarfs by $z=0$.  Looking at $z=0$ dwarf galaxy satellites with masses greater than $10^8$ M$_{\astrosun}$, we find that approximately 65$\%$ of these halos are composed of cross-polluting progenitors which have since merged into one halo.  Thus, a majority of the dwarf galaxy progenitors that cross-pollute each other at early times eventually merge to form one halo with a single, stellar population that can then be observed in the present.  These will thus become one observable chemical system, which conceals the effects of the cross-pollution.   That is, only a minority of $z=0$ dwarf galaxies would share a chemical evolution history and thus possess non-unique chemical signatures.

The analysis presented above was for an individual main halo but it was also performed for three additional Milky Way-like halos from the same simulation.  For all three, the main halo is less massive than the one described in detail above, and consequently, their satellite populations are smaller in number and also less massive.  The general cross-pollution trends shown in Figures \ref{overlap_histo.fig} and \ref{volume_overlap.fig} are reproduced almost exactly.  However, the lower mass satellites also have lower mass progenitors at $z=10$.  As a result, their calculated bubble radii are smaller.  Halos that are then still close enough to cross-pollute have a higher probability of eventually merging into a single halo.  For these three main halos, approximately 80$\%$ of their luminous satellites are comprised of cross-polluting halos which have since merged, as opposed to 65$\%$ for the more massive Milky Way-like halo. 

In summary, we conclude that main halo progenitors cross-pollute each other more frequently and more completely in terms of volume whereas dwarf galaxy progenitors tend to be much more isolated objects, both from the main halo and from one another.  Furthermore, the majority of dwarf progenitors that are cross-polluting each other at $z=10$ eventually merge into a single halo at $z=0$.  While the actual extent and significance of cross-pollution do vary among different numerical realizations, the differences are not extreme.  The overall trends found concerning preferential isolation of dwarf galaxy progenitors appears robust.

\section{Conclusion}
This study examines how the hierarchical nature of structure formation influences the  histories of objects that survive today as satellites of a larger galaxy compared to those that are accreted and destroyed during the same galaxy's formation.
In particular, it looks at whether there could be any systematic differences in spatial location and environment between satellite and halo progenitors at early times, and what these differences might mean for low-metallicity stellar populations in these systems today.

Analysis of a cosmological N-body simulation of structure formation demonstrates that, at $z=10$:
\begin{itemize}
\item main halo progenitor particles lie clustered together while dwarf galaxy progenitor particles are found on the outskirts of this cluster.
\item supernova-driven winds tend to cross-pollute main halo progenitors more than dwarf galaxy progenitors with a higher fraction of dwarf galaxy progenitors remaining chemically isolated than their main halo counterparts. 
\item a majority of dwarf galaxy progenitors that are cross-polluting at $z=10$ eventually merge to form a single halo at $z=0$.
\end{itemize}

Previous work predicting detailed abundance patterns which used semi-analytical and statistical techniques have typically assumed chemical isolation of their stellar populations \citep{Karlsson_chem,Leaman_chem,Duane_paper}.  Dwarf galaxies appear to satisfy this assumption due to their spatial separation from surrounding halos.  However, higher levels of cross-pollution for main halo progenitors and thus stellar halo stars suggest that this external source of metallicity could affect the measured abundance patterns and should be included in these types of models.

These results also have important implications for future chemical abundance studies.  For example, recently \citet{Brown_ufdcmd} have shown that the UFDs contain what are effectively single age stellar populations that are at least as old as approximately 13.7 Gyr, the age of the ancient globular cluster M92.  These primeval populations are likely good tracers of early chemical enrichment or even of a single early or primordial supernova.  Moreover, if they are chemically isolated from one another as well as the main halo, they are each a unique laboratory for chemical evolution and SN yield tests.  Thus, the Local Group  overall can be viewed as having hundreds of such independent chemical histories, instead of one or two common enrichment histories dominated by the most massive halos.  Furthermore, the independent nature of the early enrichment of different local systems suggests that each could potentially have distinct chemical signatures.  

Overall, we conclude that the isolation of dwarf galaxy progenitors can be appealed to as an explanation for current observations of abundance distribution differences between low metallicity stars in satellite galaxies and in the halo. Conversely, as data sets become more extensive,  abundance differences might be used to tell us about the progress of metal enrichment  on Local Group scale in the early Universe.  

\acknowledgements
The authors gratefully acknowledge useful conversations with Zoltan Haiman. LC and KVJ were supported in part by NSF grants AST-0806558 and AST-1107373.

\bibliographystyle{apj}     
\bibliography{library,library_kvj}

\end{document}